\documentclass[aps,prl,twocolumn,groupedaddress]{revtex4} % preprint
\usepackage{graphicx}
\usepackage[utf8]{inputenc}
\usepackage[T1]{fontenc}
\usepackage{amsmath}
\usepackage{amsfonts}
\usepackage{color} % this is just for the \todo command, remove this line from the final draft

\begin{document}

\title{Interaction-induced negative mobility: realization  in a system of two coupled Josephson junctions}

\author{M.  Januszewski}

%\author{M.  Kostur}

\author{J.  {\L}uczka}

\affiliation{Institute of Physics, University of Silesia, Katowice, Poland}

%\date{\today}

\begin{abstract} An overdamped dynamical system, biased by an external
constant force, does not exhibit negative mobility. However, when the system is coupled
to  its copy, negative mobility can arise. We show it by the example of an
experimentally realizable system   of two  coupled  resistively shunted
Josephson junctions.  
%The system is externally  shunted by  the resistance.  
The first junction is dc-biased  by a constant current  and ac-driven by an
%unbiased oscillating current of one harmonic component. 
 unbiased harmonic current.   In  the proposed  setup  one  can  control
 the transport properties of the second junction by manipulating the first
 junction only.   We  demonstrate  that 
%within the tailored   parameter  regimes,  
the second junction can exhibit both absolute negative  resistance (ANR) near
zero bias and negative resistance in the  non-linear regimes (NNR).  These  anomalous
transport features  occur  for  restricted  windows of the coupling
constant.   \end{abstract}


\maketitle

Systems under non-equilibrium conditions can display new features as well as
unexpected phenomena and processes  which are forbidden in equilibrium  systems
by fundamental laws.  One of the most prominent examples include the phenomena of
negative mobility (conductance, resistance) \cite{hop}:  when a
constant force is applied to a mobile particle, it moves in the direction
opposite to that of the force. This phenomenon is impossible in equilibrium states because it would violate
the second law of thermodynamics.  A simple  model is formulated in terms of
the Newton equation describing the one-dimensional motion of a classical Brownian
particle. In the minimal model, the particle moves in a symmetric spatially
periodic potential, is biased by a static force, and driven by  the  unbiased
harmonic force $A  \cos(\omega t)$
\cite{PhysRevLett.98.040601,PhysRevE.76.051110,PhysRevB.77.104509}. This system
is out of equilibrium and displays both absolute  negative mobility (ANM)
around zero static applied force (the linear response regime)  and negative
mobility in the non-linear response regime (NNM).   In  the linear response
regime, the long-time averaged particle velocity   tends to zero when the
static force  tends to zero. In the nonlinear response regime, the particle
velocity  can tend to zero even if the static force is far from zero.  It is
known that the  corresponding {\it overdamped}  dynamical system does not
exhibit negative mobility: the inertial  term in the Newton equation is
absolutely necessary for the negative mobility to occur.  However, phenomena
which are absent  in a single element  perhaps can occur in a system of
coupled elements.  The physics of many-body systems provides hundreds of examples.
Therefore, we are going to check whether it is possible to construct a system of two coupled overdamped particles that
exhibits negative mobility.   We  can speculate  that the combined effect
of  the time-dependent driving and the interaction between the two particles can
radically modify a one-particle overdamped dynamics yielding transport
anomalies like ANM or NNM.  We  apply the static and  time-periodic forces to
only one (say the first) particle and observe  the response of  the second
particle. In a more general context, one can manipulate  the first
particle and  by this means  control the transport properties of the second particle.  As a
possible realization of such a   system we exploit two Josephson junctions
which play the role of the two particles in the mechanical  framework. 

%%%%%%%%%%%%%%%%%%%%%%%%%%%%%%%%%%%%%%%%%%
\begin{figure}[bp]
	\centering
	\includegraphics[width=0.4\textwidth]{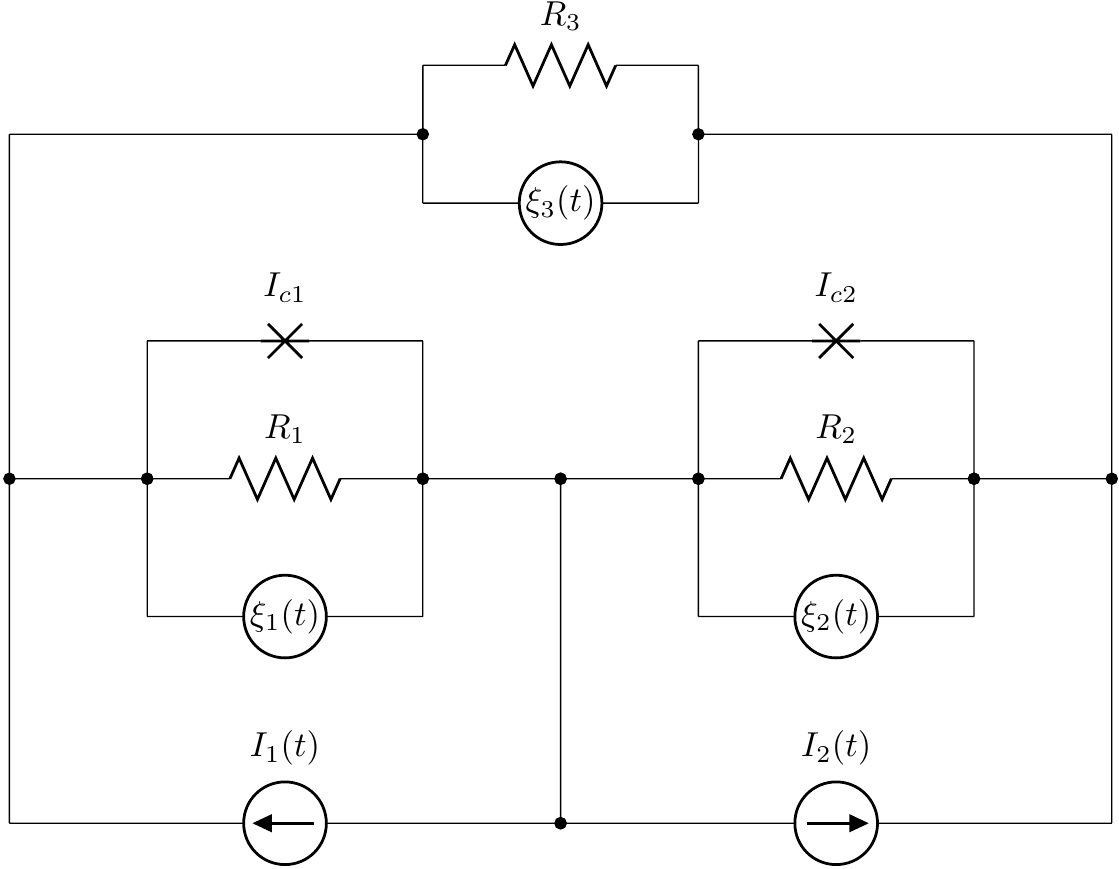}
	\caption{The system of two resistively shunted  Josephson junctions
	coupled  by  an  external  shunt resistance $R_3$ and driven by the
	currents $I_1(t)=I_1+a_1 \cos(\omega t)$ and
	$I_2(t)=I_2+ a_2 \cos (\omega t)$.}
	\label{2jj}
\end{figure}
%%%%%%%%%%%%%%%%%%%%%%%%%%%%%%%%%%%%%%%%%%%%%%%%%

%%%%%%%%%%%%%%%%%%%%%%%%%%%%%%%%%%%%%%%%%%%%
{\it Model.}--In a more general case, we can  consider  a  system consisting
of two coupled  resistively shunted Josephson junctions with the critical
currents $(I_{c1},  I_{c2})$,  resistances $(R_1, R_2)$ and  phases $(\phi_1,
\phi_2)$, respectively.  The system is externally  shunted by  the resistance
$R_3$ as shown in Fig.~\ref{2jj}. 
Moreover, the junctions can be dc-biased by
the currents $I_1$ and  $I_2$,  and ac-driven  by the currents  $a_1 \cos
(\omega t)$  and $a_2 \cos (\omega t)$, respectively.   We consider the small
junction area limit and  a regime where photon-assisted tunneling phenomena do
not contribute. The phase dynamics of the  junctions is determined by two equations,  the
dimensionless form  of which read  \cite{PhysRevB.21.118}
\begin{subequations}
	\label{1}
	\begin{align}
		\dot{\phi}_1 &= I_1 - I_{c1} \sin \phi_1 +  b_1 \cos(\omega t)  \nonumber\\
			& + \alpha \left( I_2 - I_{c2} \sin \phi_2 \right ) + \sqrt{D}\; \xi'(t),  \\
		\dot{\phi}_2 &=  \alpha \beta \left( I_2 - I_{c2} \sin \phi_2 \right) + b_2 \cos(\omega t) \nonumber \\
			&+\alpha \left( I_1 - I_{c1} \sin \phi_1 \right) + \sqrt{\alpha \beta D} \;\xi''(t),
	\end{align}
\end{subequations}
where  the dot denotes a time derivative, and the parameters  $\beta = 1 + R_3/R_1$, $b_1 = a_1 + \alpha a_2$ and  $b_2 =
\alpha (a_1 + \beta a_2)$.  The coupling  parameter  $\alpha =
R_2 / (R_2 + R_3) \in [0, 1]$ can be changed by adjusting the external resistance $R_3$.
We use the same  dimensionless units as Nerenberg
et al.~\cite{PhysRevB.21.118}:  the current unit is the averaged
critical supercurrent $\bar{I_c} = (I_{c1} + I_{c2}) / 2$, and the time
unit is $\hbar / 2e V_0$, where $V_0 = \bar{I_c} R_1 \left( R_2 + R_3 \right) / ( R_1 + R_2 + R_3 )$
is the characteristic voltage. We assume that
all resistors are  at the same temperature $T$, and that the noise sources (the
Johnson thermal noise) are represented  by zero-mean white noises $\xi_i(t)
\, (i =1, 2, 3)$   which are delta-correlated, i.e.  $\langle \xi_i(t) \xi_j(s)
\rangle = \delta_{ij} \delta(t-s)$  for $i, j \in \{1, 2, 3\}$.  The
noises $\xi'(t)$ and $\xi''(t)$,  which  appear in Eqs.~\eqref{1},  are linear
combinations of the noises $\xi_i(t), i = 1, 2, 3$.  The dimensionless noise
strength is $D = 4 e k_B T / \hbar \bar{I_c}$.  Eqs.~\eqref{1} are the
extended version of the system studied by Nerenberg et
al.~\cite{PhysRevB.21.118, PhysRevB.25.1559} which now include ac driving and
noise terms. 
 
Eqs.~(\ref{1}) describe overdamped dynamics of a hypothetical mechanical  system
of two  interacting particles of coordinates $x_1=\phi_1$ and $x_2=\phi_2$,
respectively.  The phase space of the deterministic system (\ref{1}) is
three-dimensional,   namely $\{x_1=\phi_1, x_2=\phi_2,   x_3=\omega t\}$.  Three
is the minimal dimension of the phase space to display  chaotic  evolution which
is an important feature for anomalous transport to occur
\cite{PhysRevLett.98.040601,PhysRevE.76.051110,PhysRevB.77.104509}. At a non-zero
temperature, $D>0$,  the Johnson thermal fluctuations activate a diffusive
dynamics where stochastic escape events among existing attractors become
possible. Moreover, the system  can now visit any part of the phase space and
evolve within some finite time interval by closely following  any existing
orbits, either stable or unstable.

{\it Driving applied to one  junction}. -- The detailed  analysis of the full
system~(\ref{1}) will be presented elsewhere. Here, we consider a simplified system
of  {\it two identical} junctions ($R_1= R_2,\,I_{c1}=I_{c2}=1,\,\alpha\beta=1$)
and the case  when $I_2=0$ and $a_2=0$ -- that is, when only the first junction is dc- biased and ac-driven.
In this way,  we want to manipulate the first junction only and  observe
the response of the second junction. We ask under what conditions we can
control transport properties of the second junction.  To this aim, we
numerically study the  dimensionless  long-time  averaged voltage $v_2=\langle \dot
\phi_2 \rangle $ across the second junction. The corresponding asymptotic
averaged   voltage across the first junction is denoted as $ v_1 =\langle \dot
\phi_1 \rangle$. The long-time  physical voltage is then expressed as $V_i=V_0
v_i \,( i=1, 2)$.  If the problem is formulated in terms of overdamped motion
of  classical Brownian particles, the voltage  $ v_i$ translates into the
averaged  velocity  of the first or second particles, respectively, and the dc
current $I_1$ translates into an external constant  force acting only on the
first particle. The junction resistance (or equivalently conductance)
translates then into the particle mobility.  One can use this analogy to simplify
the visualization of  transport processes in junctions.  The voltage $v_i
=v_i(I_1)$ is typically a nonlinear and non-monotonic function of  the dc
current $I_1$.  In the "normal" transport regime, the voltage is positive $v_i>0$ for positive bias $I_1>0$,
i.e.  the "nonlinear resistance" or the static
resistance $R = R(I_1)= v_i(I_1)/I_1 $ at a fixed bias current  $I_1$  is
positive.  
If $R<0$,  i.e. when the system response is  opposite  to the external load,  it refers to the
anomalous transport regime with  ANR or NNR. 
%When  $R = v_i(I_1)/I_1 < 0$,  the phenomenon of the negative  resistance
%(conductance,  mobility) occurs. 
% \todo{This phenomemon should not be identified with the more well-known differential
%negative mobility.} 
% If $R<0$ holds true in the linear respose regime, i.e.
%when  $v_i = R I_1$ for $I_1 \to 0$,  it is termed  ANR in analogy to the absolute negative mobility of  Brownian particles. 

We begin the analysis of the  system (\ref{1}) by some general remarks about  its
long-time behavior.  As expected from symmetry, in the zero bias case, i.e.
when $I_1=0$,   there is no net transport. If the dc bias $I_1$  is
sufficiently  large in comparison to the amplitude $a_1$ of the ac driving, the
voltage across both junctions has the same sign as the dc current. This is rather
obvious  because the driving is not important in such a regime. More interesting
effects can take place in the regime of small $I_1$. However, the parameter space
$\{\alpha, I_1,  a_1,  \omega,  D\}$  is 5-dimensional and thus too large for
an extensive numerical scan.  Therefore,  a number of low resolution scans over the
parameters $\{\alpha, a_1, \omega\}$ at fixed values of $I_1$ and $D=0$   was
first performed. Next,  the interesting  regions of the parameter space were analyzed
in more detail and at a higher resolution.  These initial scans were done for
initial values of the phases $\phi_1$ and $\phi_2$ randomly chosen from the
interval  $[0; 2\pi]$.  They were then supplemented by scans at non-zero
temperatures.  We have found  that  for $I_1 < 0.1$ the modulus of the average
voltage across  both  junctions takes its highest values for $\omega \in (0,
1)$ and is gradually diminished for higher frequencies.  At large  ac-driving
frequencies ($\omega > 5$)  there is no noticeable net transport regardless of
the values of other parameters.

\begin{figure}
	\centering
	\includegraphics[width=0.5\textwidth]{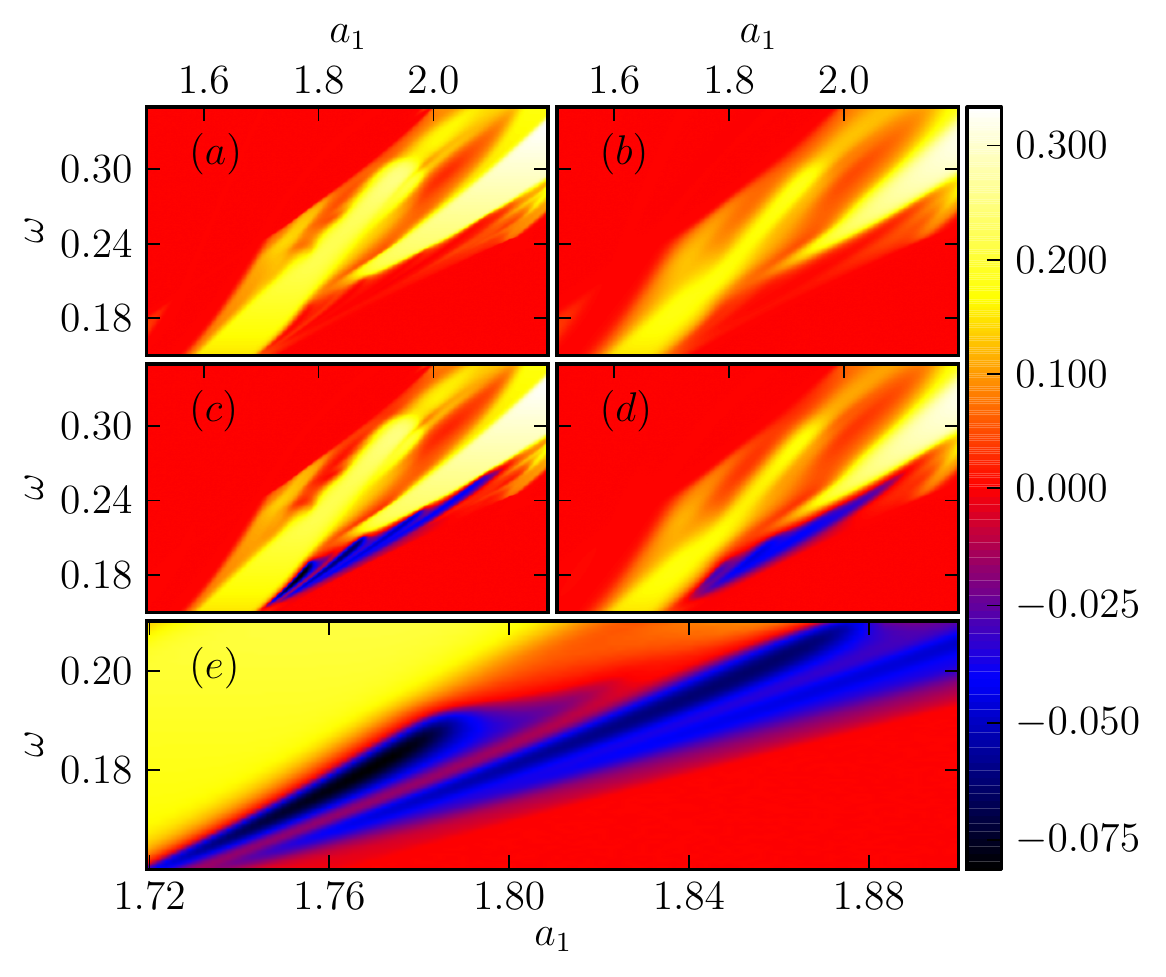}
	\caption{The transport properties of the driven system of two
	Josephson junctions in the parameter space $\{a_1, \omega\}$
	of the ac driving acting on the first junction at a
	representative coupling  value of $\alpha =0.77$, the dc bias
	$I_1 = 0.05$ and  $a_2 = I_2 = 0$.  In panels (a) and (b) the
	averaged voltage $v_1$ across the first junction is shown at
	two dimensionless temperatures  $D = 5\cdot 10^{-5}$ and
	$2.5 \cdot 10^{-4}$, respectively.  Panels (c) and (d):   the
	averaged voltage $v_2$  across the second  junction is depicted at
	the same temperatures as in   panels (a) and (b), respectively. 
Panel (e)  presents an enlarged
part of panel (c): the plot  reveals the island structure of regions of negative resistance of the second junction.}
	\label{fig:2jj_ident_anm}
\end{figure}

For  fixed but small values of the dc bias $I_1<0.1$,  strips of non-zero
average voltage are  clearly  visible  in the parameter space $\{a_1,
\omega, \alpha\}$.  For  weak coupling of two junctions (small $\alpha$), there
is no net transport in the second junction.     For stronger coupling,  strips
of non-zero average voltage start to appear at progressively lower values of
the amplitude $a_1$ of the ac driving.  The strips are also visible in the
plots of the average voltage of the first junction, which means that they
represent regimes of the parameter space where both junctions operate  in
synchrony.  E.g. if  $I_1=0.05$, for $\alpha \in [0.66; 0.9]$ and $a_1 \approx
2.0 \pm 0.5$ the structures visible in the plots become more complex and
regions of negative conductance  of the second junction  can be identified.   As an example, we illustrate in Fig.~\ref{fig:2jj_ident_anm} the emerging
asymptotic, averaged voltages across the first and the second junction    for
two  different  dimensionless temperatures $D$.  We show  how the
parameter plane $\{a_1, \omega\}$ is divided into regions of normal ($v_1(I_1)
> 0$  and $v_2(I_1) > 0$  for $I_1>0$ ) and anomalous ($v_2(I_1) < 0$  for $I_1
>0$) transport.  In doing so here we  do not discriminate between ANM and NNM.
Both these transport behaviors are jointly presented. A
more careful investigation of the parameter regimes,   where negative mobility
appears,   reveals a rich structure  which is gradually smeared out  by
increasing temperature. 
\begin{figure}
	\centering
	\includegraphics[width=0.45\textwidth]{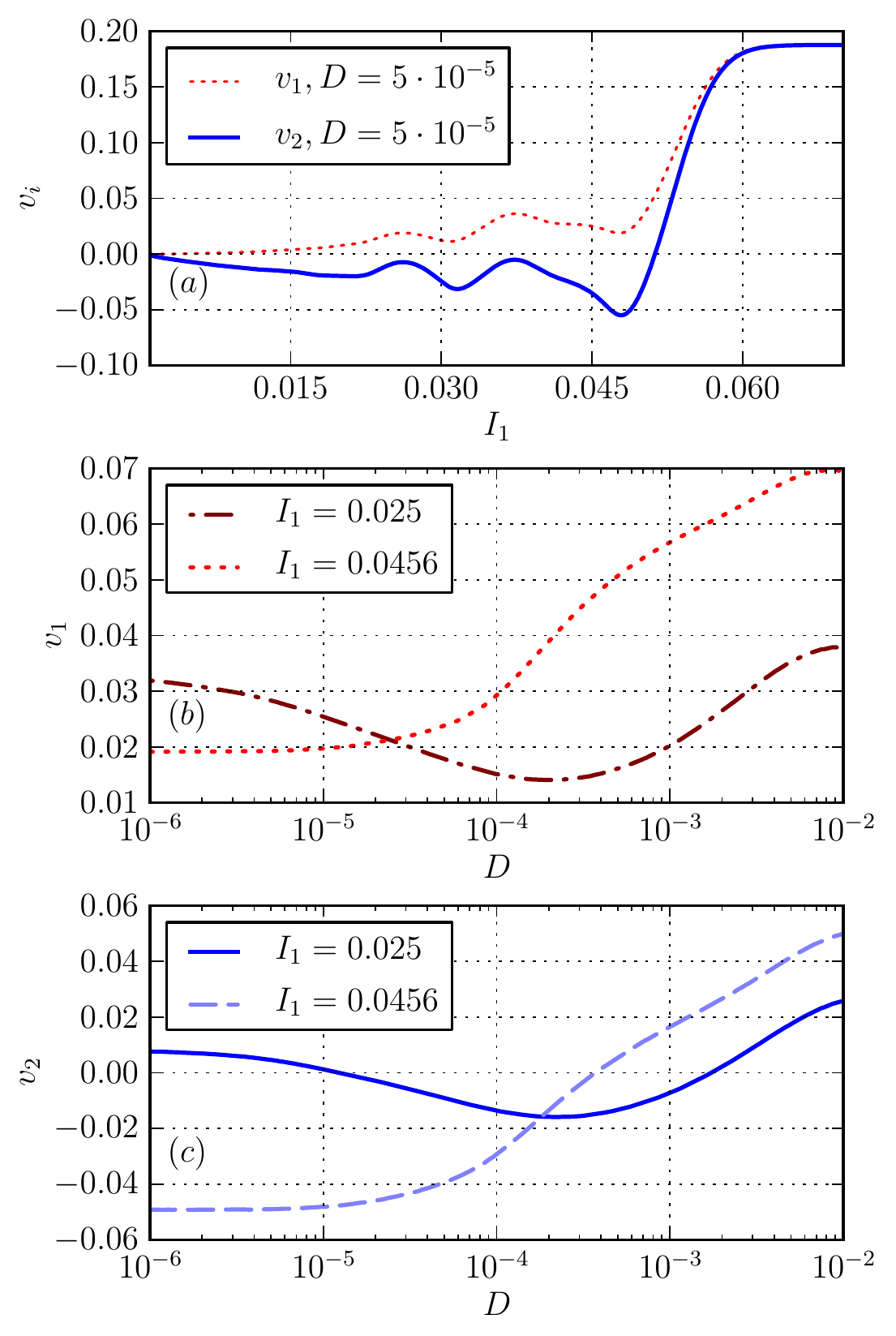}
	\caption{The long-time averaged voltages $v_1$ and $v_2$  of the first and second junctions,
		respectively.  In panel (a) the dependence on the dc bias is depicted  at a fixed
		temperature $D=5 \cdot 10^{-5}$.  Panels (b) and (c) depict the temperature dependence of
		$v_1$ and $v_2$, respectively. In panel (c), please note two distinct
		mechanisms for the negative voltage of the second junction:  noise-induced  for
		$I_1=0.025$ and deterministic,  chaotic-assisted  for $I_1=0.456$.  The parameters
		are:  the coupling strength $\alpha = 0.77$,  amplitude  $a_1 = 1.7754$ and
		frequency $\omega = 0.1876$ of the ac driving, $a_2 = I_2 = 0.0$.}
	\label{fig:2jj_anm_prof1}
\end{figure}
We will now discuss the  dependence of the voltage on the second junction of the dc  bias
$I_1 >0$ acting on the first junction, in the regime of anomalous transport
illustrated in Fig.~\ref{fig:2jj_ident_anm}.  We fix both the amplitude $a_1$ and
the frequency $\omega$ of the ac driving acting on the first junction.  The result
is presented in Fig.~\ref{fig:2jj_anm_prof1}.  Most prominent are
the intervals of $I_1$ where the voltage $v_2$ is
negative.  Two anomalous effects are detected:  ANR  for $I_1 \to 0$ (the linear response regime)  and NNR  when $I_1$  has a value  remote  from zero.
In the case presented in Fig.~\ref{fig:2jj_anm_prof1}(a),  one can observe three intervals of
the bias $I_1$  where NNR occurs.  This is to be contrasted with the averaged
voltage of the first junction which is never negative.  However, one can note
a rough  synchronization in the dependency of $v_1$ and $v_2$  on the bias $I_1$:
simultaneous increases  and  decreases of both voltages in some intervals of the bias.
Similar synchronization is also observed in the temperature dependence, see Fig.~\ref{fig:2jj_anm_prof1},
panels (b) and (c).   A closer inspection of the panel (c) of
Fig.~\ref{fig:2jj_anm_prof1} reveals
another interesting  result:   there are two  fundamentally  different mechanisms
generating   anomalous transport.  In the case $I_1=0.025$, the negative
resistance  is induced by thermal fluctuations.  There is a  finite interval of
temperature where this effect occurs. For very low temperature (lower than $1.2
\cdot 10^{-5}$) and high temperature (greater  than $2 \cdot 10^{-3}$) the
voltage   $v_2$  takes positive  values.   Between these two temperatures, the
voltage is negative.  In contrast, in the case $I_1=0.0456$, the negative
resistance  is  generated by purely deterministic dynamics.  Even for
zero-temperature $D=0$, the resistance is negative.  For this chaos-assisted
mechanism, temperature plays  a destructive role: increasing temperature
monotonically diminishes the negative voltage $v_2$,  and after crossing zero
at  some critical temperature,  the voltage  assumes  positive values.

\begin{figure}
	\centering
	\includegraphics[width=0.5\textwidth]{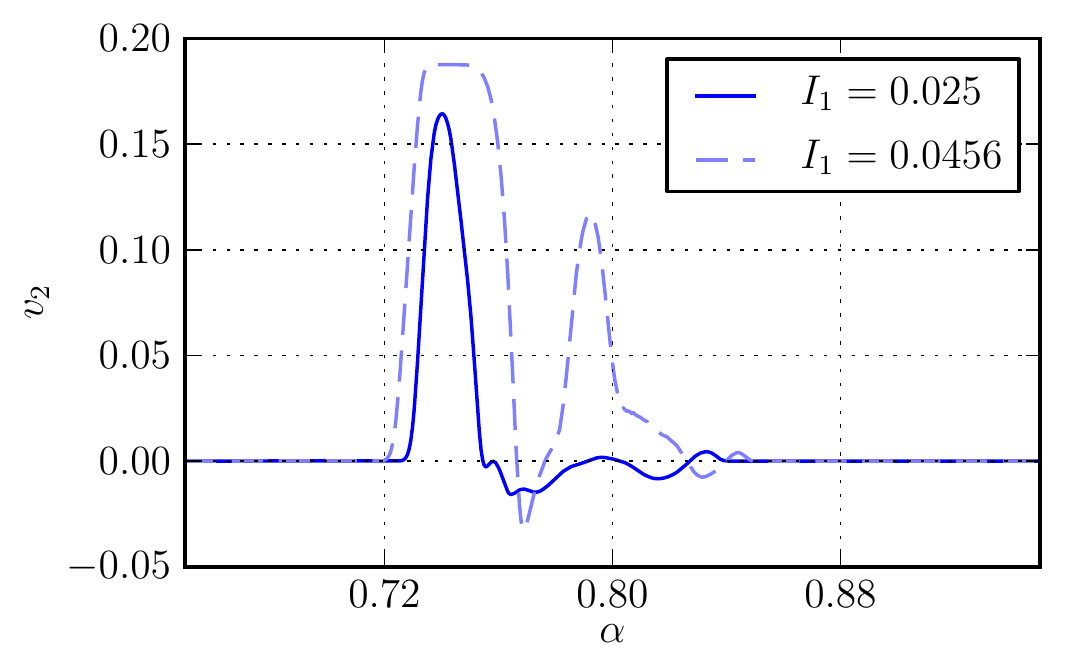}
	\caption{The influence of  the coupling parameter $\alpha$ on transport properties of
		the second junction. The  long-time averaged voltage  $v_2$ across the
		junction is depicted for two fixed values of  the  externally applied
		dc current $I_1$ and $D=5 \cdot 10^{-5}$.  For $I_1=0.025$ there are three intervals of $\alpha$
		where the voltage is negative,  while for $I_1=0.0456$ there are two intervals
		of such  $\alpha$.  Other parameters are the same as  in Fig.~\ref{fig:2jj_anm_prof1}.}
	\label{fig:anm_aprof}
\end{figure}

It should be emphasized that the anomalous transport effects discussed above
are all caused by the coupling between two junctions -- without coupling,  the
negative voltage vanishes.  Fig.~\ref{fig:anm_aprof} shows how the average voltage of the second
junction  depends on the coupling constant $\alpha$. We can see that there are
always finite windows of $\alpha$ for which this effect can be observed. The
location and  size of these windows depend on values of other system
parameters.  For the regimes depicted in Fig.~\ref{fig:2jj_anm_prof1}, we illustrate it in Fig.~\ref{fig:anm_aprof}. In
the case when the bias  $I_1=0.025$, up to coupling  strength around
$\alpha=0.73$ the average voltage  is zero valued (numerically speaking the
current is negligibly small).  Starting from the mentioned value the transport
settles in and later the voltage reaches the highest value of $v_2 \approx
0.18$ at $\alpha \approx 0.741$. Next, the voltage  stays positive up to $\alpha  \approx 0.753$
and  then changes its sign for the first time,
resulting in an anomalous  response -- the voltage assumes negative values.
This situation holds up to $\alpha \approx 0.758$ and then the current turns into
the positive  but very small values. The same scenario is repeated  two more times 
 -- between the values of $\alpha  \approx (0.76,  0.788)$ and $(0.806,
0.826)$. In the case $I_1=0.0456$, there are two distinct intervals  of the
coupling constant for which the voltage  $v_2$ is negative.

Adding a constant bias to the second particle has a destructive
impact on the negative mobility -- the higher the value of $I_2$, the smaller
the parameter area where negative mobility can be observed. For instance, we find
$0.1$ to be the limiting value of $I_2$ at which areas of negative mobility
cease to exist everywhere in the analyzed area when $I_1 = 0.05$.  Increasing
$I_2$ has a general smoothing effect on the transport properties of the second
particle -- at higher values of $I_2$, the fine details visible for $I_2=0$
gradually disappear.  Increasing $I_2$ also causes all features to move
slightly towards higher values of $a_1$ and $\omega$.  It is therefore possible
to find areas of the parameter space where $I_2$ induces a negative mobility
effect  as well as areas where it increases the strength of an existing
negative mobility effect.\\
It is of great importance to construct simple models which allow to explain and
clarify the understanding of  unusual and anomalous transport properties not
only in physical but also in biological systems such as  e.g. the bi-directionality of
the net cargo transport inside  living cells \cite{biol}. Such models  can also serve as
a  basis for  the construction of  more realistic and quantitative  models for
transport of  interacting carriers  in collective  systems.  In this paper, we
constructed a simple model of two coupled elements, described in terms of
overdamped dynamics, which exhibits negative mobility both in the linear and
non-linear regimes.  There are regimes where both particles can be transported
in the same direction as the external static force  acting on the first
particle. There are also  regimes where only the second particle can be
transported in the opposite direction to bias.  The interaction between two
particles  constitutes a crucial ingredient of the model: without coupling, the
second particle is not transported at all, while the first  particle can be
transported only in the direction of the dc bias.  The system  can be
manipulated in other ways.  The source of energy, the driving $I_1(t)=I_1+a_1
\cos (\omega t)$, can be replaced by another  source of energy such as e.g.  an unbiased
multi-harmonic force or, in biological systems, chemical reactions. An open
question is what minimal conditions should be satisfied to generate  anomalous
transport, in particular bidirectional motion of two particles.  Our conjecture
is that chaotic dynamics is necessary, but not sufficient.  Finally,
it is a promising topic  which can stimulate  experimentalists  to perform
measurements  testing our findings in systems of two coupled Josephson
junctions where coupling can be precisely controlled by the external resistance
and,  moreover,  to plan experiments for other systems.

\end{document}